\documentclass[aps,prl,twocolumn,amssymb,amsmath,longbibliography]{revtex4-2}

\usepackage[pdftex]{graphicx}		
\usepackage{bm}
\usepackage{array}
\usepackage{amsmath}
\usepackage{txfonts}
\usepackage{color}

\begin{document}


\title{Sample dependence of the half-integer quantized thermal Hall effect in the Kitaev candidate $\alpha$-RuCl$_3$}

\author{M. Yamashita$^1$}
 \email[]{my@issp.u-tokyo.ac.jp}
\author{J. Gouchi$^1$}
\author{Y. Uwatoko$^1$}
\author{N. Kurita$^2$}
\author{H. Tanaka$^2$}

\affiliation{$^1$The Institute for Solid State Physics, The University of Tokyo, Kashiwa, 277-8581, Japan}
\affiliation{$^2$Department of Physics, Tokyo Institute of Technology, Tokyo 152-8551, Japan}

\date{\today}

\begin{abstract}
We have investigated the sample dependence of the half-integer thermal Hall effect in $\alpha$-RuCl$_3$ under a magnetic field tilted 45 degree from the $c$ axis to the $a$ axis. We find that the sample with the largest longitudinal thermal conductivity ($\kappa_{xx}$) shows the half-integer quantized thermal Hall effect expected in the Kitaev model. On the other hand, the quantized thermal Hall effect was not observed in the samples with smaller $\kappa_{xx}$. We suggest that suppressing the magnetic scattering effects on the phonon thermal conduction, which broaden the field-induced gap protecting the chiral edge current of the Majorana fermions, is important to observe the quantized thermal Hall effect.
\end{abstract}

\maketitle

Non-trivial topology in a condensed-matter state realizes a quantization of a physical quantity. One of the most fundamental examples is the quantized Hall conductivity in a quantum Hall system, where the quantized Hall conductivity is given by the Chern number determined by the topology of the system~\cite{Klitzing1986}.

A new intriguing case of this topological quantization is a Kitaev magnet~\cite{Kitaev2006, Motome2020}. In the Kitaev model, localized spin-1/2 moments on a two-dimensional (2D) honeycomb structure are coupled each other by bond-dependent Ising interactions. The frustration effect of this Kitaev Hamiltonian prevents the spins to order even at the zero temperature, realizing a quantum spin liquid state. Remarkably, this ground state of the Kitaev Hamiltonian is exactly solvable. The ground state has been shown to be characterized by the two kinds of elementary excitations; itinerant Majorana fermions and localized $Z_2$ fluxes. In a magnetic field, this itinerant Majorana fermions have topologically non-trivial gapped bands with the Chern number $C = \pm 1$, giving rise to a quantized chiral edge current. In contrast to a quantized chiral edge current of electrons in a quantum Hall system, this chiral edge current is carried by the charge neutral Majorana fermions. Therefore, this quantized chiral edge current has been predicted to appear in the 2D thermal Hall conductivity as
$\kappa_{xy}^{2D}/T = (C/2) q_t$, where $q_t = (\pi/6) k_B^2/\hbar$.

Materializing the Kitaev model has been suggested in several Mott insulators with a strong spin-orbit coupling~\cite{JackeliKhaliullin2009}. One of the most studied Kitaev candidates is $\alpha$-RuCl$_3$ in which a 2D honeycomb structure of edge-sharing RuCl$_6$ octahedra has been shown to have a dominant Kitaev interaction~\cite{Plumb2014,Heung-SikKim2015,Winter2017}. 
Various measurements~\cite{Do2017,Sandilands2015,Nasu2016,Banerjee2016,Banerjee2017,KasaharaP2018} have reported Kitaev-like signatures above the antiferromagnetic (AFM) ordering temperature of $T_{\textrm N} \sim 7$\,K (Refs.\,\cite{Kubota2015, Sears2015, Cao2016, Do2017}).
This magnetic order can be suppressed by applying a magnetic field of $\sim 8$\,T in the $a$--$b$ plane~\cite{Kubota2015,Johnson2015,Sears2017}, enabling one to study the Kitaev QSL down to lower temperatures. Most remarkably, thermal Hall measurements done under an in-plane field have shown the half-integer quantized thermal Hall conduction~\cite{KasaharaN2018,Yokoi2020}, indicating the presence of a chiral edge current of the Majorana fermions protected by the field-induced gap. However, details of this field-induced gap are unknown because the Kitaev Hamiltonian loses its exact solvability in a magnetic field.

It has been reported that this quantized thermal Hall effect has a sample dependence associated with the longitudinal thermal conductivity ($\kappa_{xx}$)~\cite{Yokoi2020}.
This $\kappa_{xx}$ dependence may imply a scattering effect on the field-induced gap protecting the chiral edge current. A similar scattering effect has been discussed in the intrinsic anomalous Hall effect (AHE) in ferromagnetic metals~\cite{Shiomi2010} where a broadening of the gap by scattering effects is suggested to destroy the intrinsic AHE in a less conductive metal. Therefore, further studies of the $\kappa_{xx}$ dependence of this quantized thermal Hall effect may provide information with respect to the unknown field-induced gap.
It is also important to confirm the reproducibility  of the quantized thermal Hall effect.

In this Letter, we report the sample dependence of the longitudinal ($\kappa_{xx}$) and transverse ($\kappa_{xy}$) thermal conductivity of three single crystals of $\alpha$-RuCl$_3$. 
We confirm the reproducibility of the half-integer quantized thermal Hall effect in a sample showing the largest $\kappa_{xx}$ among the three crystals.
On the other hand, the other samples with smaller $\kappa_{xx}$ show $\kappa_{xy}$ much smaller than the value expected for the quantization. We also find that a sample with a larger $\kappa_{xx}$ shows a larger decrease of the magnetic susceptibility below $T_{\textrm N}$, in addition to a larger field-increase effect of $\kappa_{xx}$, showing that magnetic scattering effects are more strongly suppressed by magnetic fields in a sample with a better quality. From these results, we suggest that suppressing this magnetic scattering effect plays an important role to realize the quantized thermal Hall effect.

\begin{figure*}[!tbh]
	\centering
	\includegraphics[width=\linewidth]{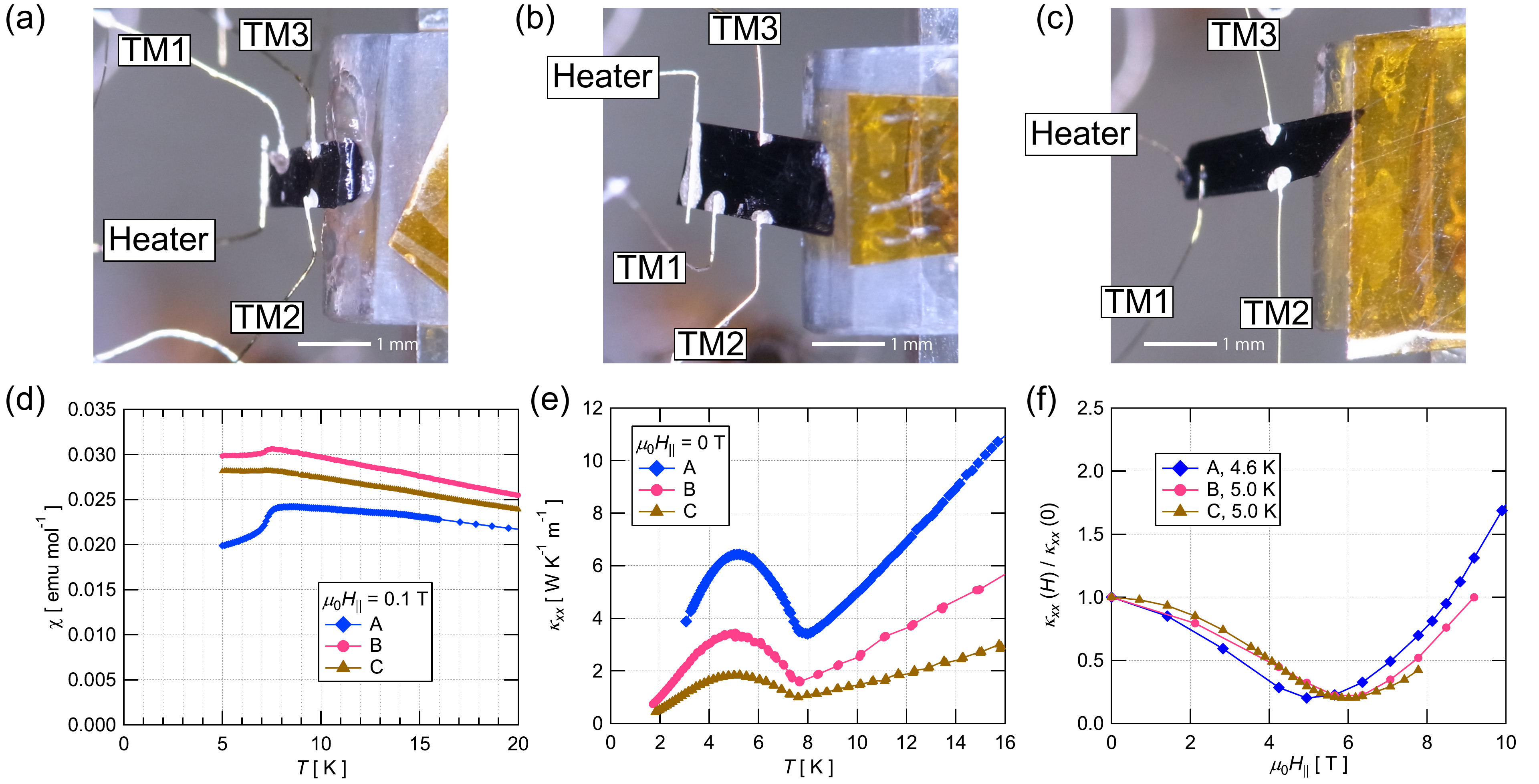}
	\caption{
		(a--c) Setup pictures of the thermal conductivity and the thermal Hall measurements of sample A, B and C, respectively. Note that these pictures were took at different angles to the crystal.
		The thermal connections to the heater and the three thermometers (TM1--3) are indicated in each picture. 
		The heat current was applied along the $a$ axis (the long axis of the crystal) and the magnetic field was applied 45 degree from the $c$-axis to the $a$-axis.
		(d) The temperature dependence of the magnetic susceptibility at 0.1\,T applied along  the $a$-axis. 
		(e) The temperature dependence of the longitudinal thermal conductivity $\kappa_{xx}$ at zero field. 
		(f) The field dependence of $\kappa_{xx}$ at 4.6\,K (sample A) and at 5.0\,K (sample B and C). The vertical axis is normalized by the zero-field value $\kappa_{xx}(0)$. The horizontal axis shows the in-plane field $\mu_0 H_{\parallel} = \mu_0 H / \sqrt{2}$.
	}
	\label{chi_and_kxx}
\end{figure*}

Single crystals used in this work were synthesized by a Bridgeman method as described in Ref.~\cite{Kubota2015}. We have measured both $\kappa_{xx}$ and $\kappa_{xy}$ of three single crystals (sample A--C) of $\alpha$-RuCl$_3$. A typical sample size was 2.5\,mm {$\times$} 1.0\,mm {$\times$} 0.03\,mm. These thermal-transport measurements were done by using a one-heater-three-thermometers method (see Figs. (a)--(c)) as described in Ref.~\cite{YamashitaJPCM2019}. The measurement cell was the same with that used for the previous work (sample 2 in Ref.~\cite{KasaharaP2018}). A heat current was applied along the $a$ axis of the sample, and a magnetic field $H$ was applied 45 degree from the $c$-axis to the $a$-axis. We denote the in-plane field $\mu_0 H_{\parallel}$ as $\mu_0 H_{\parallel} = \mu_0 H / \sqrt{2}$.
We note that, because $\alpha$-RuCl$_3$ is easily deformed owing to the van der Waals structure,  it is important to carefully handle $\alpha$-RuCl$_3$ to avoid applying stresses on the sample. However, some degree of stress is unavoidable especially when taking out a single crystal from the chunk of the crystals obtained by the Bridgeman method. Therefore, we checked the sample quality by the magnetic susceptibility and $\kappa_{xx}$ measurements prior to $\kappa_{xy}$ measurements.

The temperature dependence of the magnetic susceptibility ($\chi$) was checked for all samples prior to the thermal conductivity measurements (Fig.\,1(d)). 
To avoid the effect of the in-plane anisotropy of $\chi$~\cite{Lampen-Kelley2018}, the magnetic field was applied along the $a$-axis for all the $\chi$ measurements.	
As shown in Fig.\,1(d), no anomaly is observed at 14\,K, showing the absence of the additional magnetic transition caused by stacking faults~\cite{Kubota2015,Cao2016}.
We further checked the sample quality by heat capacity measurements performed after the thermal-transport measurements. We confirmed that the anomaly by the additional magnetic transition is absent or small for all samples~\cite{SM}.
The AFM transition at $T_{\textrm N} \sim 8$\,K is clearly seen in the all samples. The largest decrease of $\chi(T)$ below $T_{\textrm N}$ is observed in sample A. This decrease is smaller in sample B and the smallest in sample C.

Figure 1(e) shows the temperature dependence of $\kappa_{xx}$ at zero field. As shown in Fig. 1(e), $\kappa_{xx}$ of all samples shows a very similar temperature dependence with that of previous works~\cite{Hentrich2018,KasaharaP2018,KasaharaN2018,Hentrich2019,Yokoi2020}. The magnetic transition to the AFM phase is clearly seen by the onset of the increase of $\kappa_{xx}$ below $T_{\textrm N}$. On the other hand, the magnitude of $\kappa_{xx}$ is very different for each sample; $\kappa_{xx}$ of sample A is the largest among the samples, which is ~4 times larger than that of sample C. 
A difference of the magnitude of $\kappa_{xx}$ in the same compounds is given by the difference in the mean free path of the heat carriers~\cite{Berman} which reflects the scattering strength on the carriers.
Therefore, this large sample dependence in the magnitude of $\kappa_{xx}$, comparing to that in $\chi$, shows a high sensitivity of $\kappa_{xx}$ measurements on the sample quality. 
Also, this sample dependence of $\kappa_{xx}$ well correlates to that of the decrease of $\chi$ below $T_{\textrm N}$. A sample with a larger decrease of $\chi$ below $T_{\textrm N}$ shows a larger $\kappa_{xx}$.

Figure 1(f) shows the field dependence of $\kappa_{xx}$ at $\sim 5$\,K. By normalizing the zero-field value of each sample, a very similar field dependence can be clearly seen. As shown in Fig.\,1(f), $\kappa_{xx}(H) / \kappa_{xx}(0)$ of all samples shows the minimum of $\kappa_{xx}$ at the in-plane field of $H_\textrm{min} = 5$--6\,T which corresponds to the critical field of the AFM phase~\cite{Kubota2015,Johnson2015,Sears2017}. Above the critical field, $\kappa_{xx}(H)$ is increased as increasing field. This increase is larger in a sample with a larger $\kappa_{xx}$.

\begin{figure}[!tbh]
	\centering
	\includegraphics[width=0.8\linewidth]{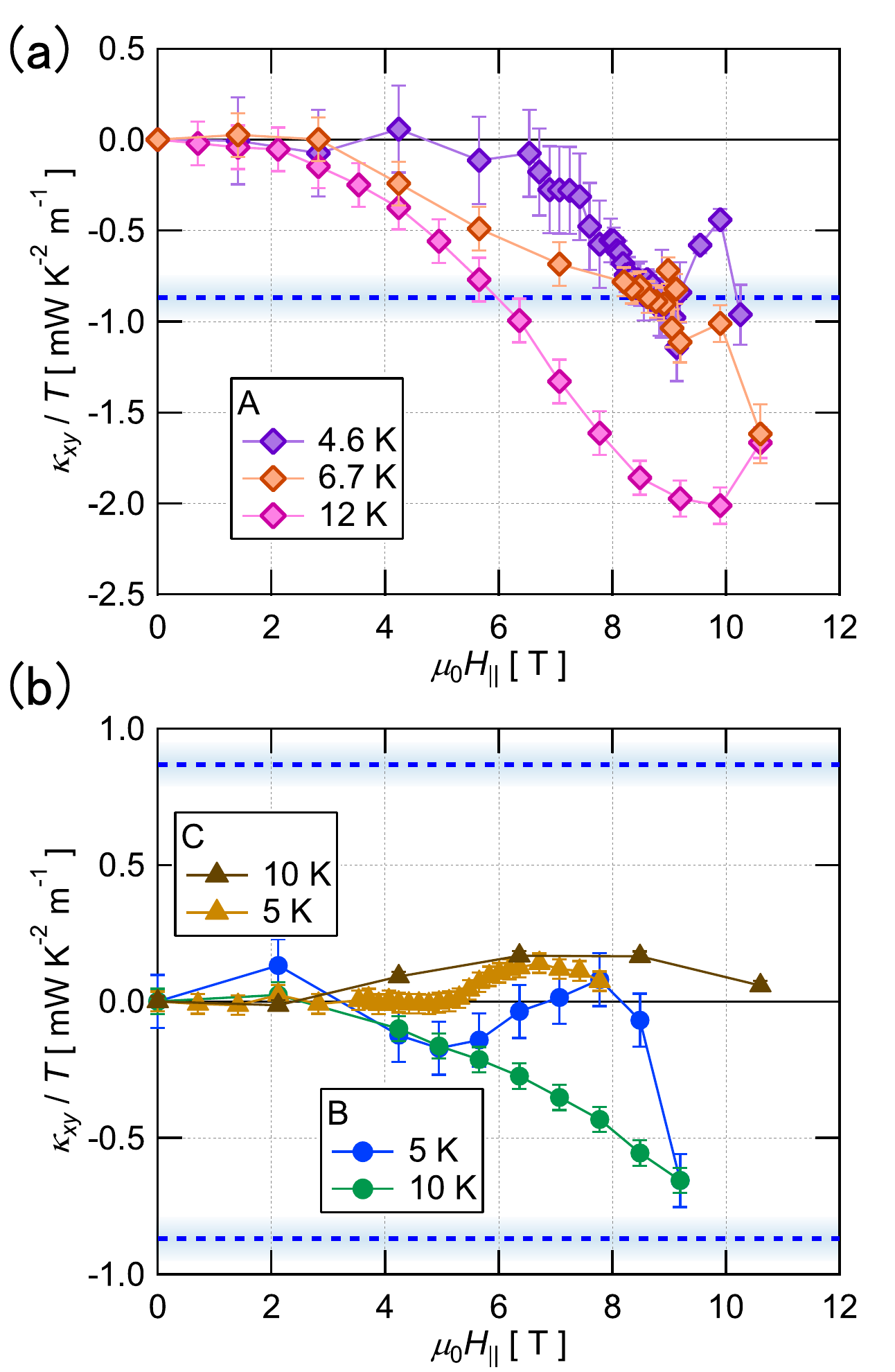}
	\caption{
		The field dependence of the thermal Hall conductivity divided by the temperature $\kappa_{xy}/T$ of sample A (a) and sample B and C (b). The dotted lines show the value corresponding to the half-integer quantized thermal Hall effect (see the main text for detail).
	}
	\label{kxy_vs_H}
\end{figure}

The field dependence of the thermal Hall conductivity at different temperatures is shown in Figs.\,\ref{kxy_vs_H}. For a comparison, the value corresponding to the half-integer quantized thermal Hall $\kappa_{xy}^{2D} / (Td )= \pm q_t /(2 d)$, where $d=0.57$\,nm is the distance between the 2D honeycomb layers of RuCl$_3$, is shown as the dotted lines.

As shown in Figs.\,\ref{kxy_vs_H}, the sign of $\kappa_{xy}/T$ at 10 K is negative in sample A and sample B whereas it is positive in sample C. This sample dependence may be related to the angle between the $a$ axis and the magnetic field, which is discussed to be negative (positive) for 45 (135) degree~\cite{Yokoi2020}. In this work, we only discuss the magnitude of $\kappa_{xy}/T$.
 
\begin{figure*}[!tbh]
	\centering
	\includegraphics[width=\linewidth]{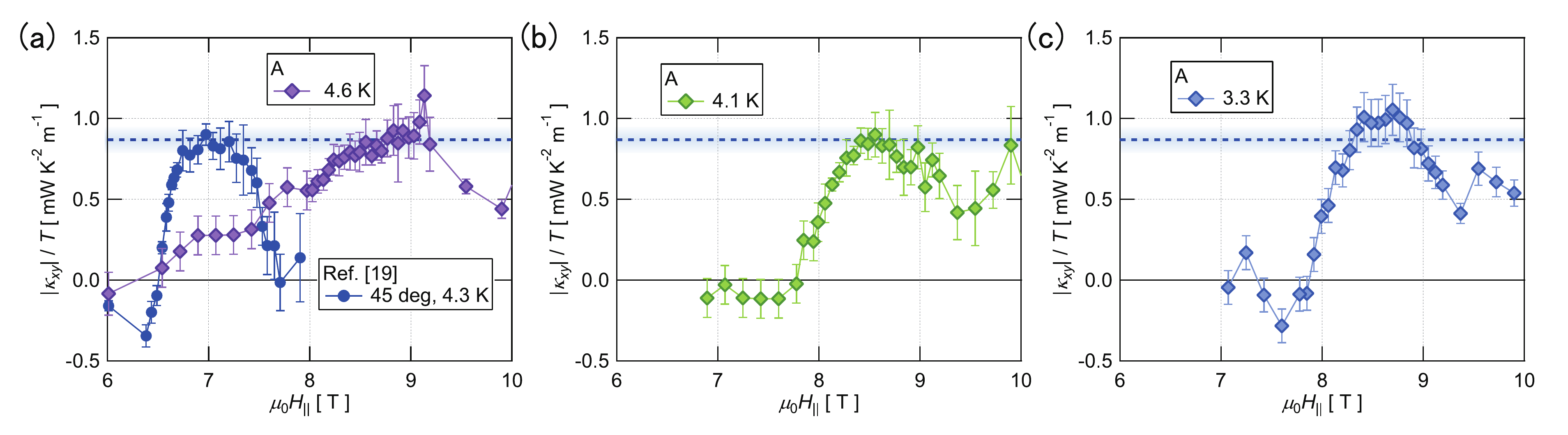}
	\caption{
		The field dependence of $|\kappa_{xy}|/T$ of sample A at 4.6 (a), 4.1 (b), and 3.3 K (c). The data in the previous report~\cite{KasaharaN2018} is also plotted in (a). The dotted lines show the value corresponding to the half-integer quantized thermal Hall effect (see the main text for detail).
	}
	\label{kxy_T_sampleA}
\end{figure*}

As shown in Fig.\,\ref{kxy_vs_H}(a), sample A shows the largest $|\kappa_{xy}|/T$. At 12 K, $|\kappa_{xy}|/T$ of sample A becomes larger than the half-integer quantized value $q_t /(2 d)$ for $H > H_\textrm{min}$. The field dependence of $\kappa_{xy}/T$ of sample A becomes flat for $\mu_0 H_{\parallel} \sim 9$\,T at lower temperatures. At the same time, the magnitude of $\kappa_{xy}/T$ at the flat region becomes close to $q_t /(2 d)$. 
On the other hand, as shown in Fig.\,\ref{kxy_vs_H}(b), the magnitudes of $\kappa_{xy}/T$ of sample B and C remain much smaller than $q_t /(2 d)$ for all temperature and field range we measured. Moreover, $\kappa_{xy}/T$ of sample B shows a very different field dependence with sign changes for $H > H_\textrm{min}$.

The field dependence of $|\kappa_{xy}|/T$ of sample A was further checked at lower temperatures (Figs.\,\ref{kxy_T_sampleA}).
As shown in Figs.\,\ref{kxy_T_sampleA}, the flat field dependence of $|\kappa_{xy}|/T$ observed for 8--9\,T persists down to 3.3\,K at $q_t /(2 d)$ within our experimental error of $\pm 10$\%.
These results demonstrate the reproducibility of the half-integer quantization of $|\kappa_{xy}|/T$ with respect to both magnetic field and temperature.
On the other hand, compared to the data in the previous report~\cite{KasaharaN2018}, the quantization of $|\kappa_{xy}|/T$ is observed at higher fields despite the similar $H_\textrm{min}$.
Quantization of $|\kappa_{xy}|/T$ at higher fields has also been reported in Ref.\,\cite{Yokoi2020}.

Here we discuss the sample dependence of $\kappa_{xx}$ and $\kappa_{xy}$.
From the previous $\kappa_{xx}$ measurements for both in-plane and out-of-plane transport~\cite{Hentrich2018}, the dominant heat carrier in $\alpha$-RuCl$_3$ has been shown as phonons.
The difference of the phonon thermal conductivity of the same compound is given by the different length of the phonon mean free path which is limited by scattering effects on phonons~\cite{Berman}.
Therefore, the different magnitudes of $\kappa_{xx}$ of different samples are determined by the different scattering strength on the phonons.
As shown in Fig.\,\ref{chi_and_kxx}(e), all samples show a very similar field dependence with a large reduction of $\kappa_{xx}$ at $H_\textrm{min}$.
This field dependence of $\kappa_{xx}$ indicates that a magnetic-field dependent scattering mechanism on phonons is dominant in all samples.
In fact, the analysis of the temperature dependence of $\kappa_{xx}$ by the Callaway model done in Ref.~\cite{Hentrich2018} has suggested that a resonant magnetic scattering is the most dominant. Therefore, the different magnitudes of $\kappa_{xx}$ in different samples are attributed to different strengths of the magnetic scatterings on phonons.

As shown in Fig.\,\ref{chi_and_kxx}(f), the increase of $\kappa_{xx}$ above $H_\textrm{min}$ is largest in sample A, and is smaller in sample B and C in order of the magnitude of $\kappa_{xx}$.
This sample dependent increase above $H_\textrm{min}$ shows that the magnetic-field dependent scattering is more strongly suppressed in a sample with a larger $\kappa_{xx}$.
In addition to this relation between the magnitude and the field dependence of $\kappa_{xx}$, a sample with a larger $\kappa_{xx}$ shows a larger decrease of $\chi(T)$ below $T_{\textrm N}$ as shown in Fig.\,\ref{chi_and_kxx}(a). This decrease of $\chi(T)$ below $T_{\textrm N}$ reflects the magnitude of the AFM order, showing that a larger decrease of $\chi(T)$ is observed in a sample with a better quality.
Therefore, a larger field-suppression on the magnetic-field dependent scattering is observed in a better-quality sample.
Given that the quantized $\kappa_{xy}$ is observed only in sample A showing the largest suppression of the magnetic-field dependent scattering, we conclude that the suppression of the magnetic-field dependent scattering in a high-quality sample is necessary to realize the quantized thermal edge current.
The different field region of the quantized thermal Hall in this work from that of the previous work~\cite{KasaharaN2018} may imply that a larger magnetic field is required to stabilize the chiral edge current in our sample.

A possible mechanism for the dissipation of the quantized thermal Hall current by the magnetic scatterings on phonons is that the field-induced gap, protecting the chiral edge current, is closed by a band broadening effect by disorders.
A similar mechanism has been put forward in the intrinsic AHE in ferromagnetic metals in which the intrinsic AHE is suggested to be dissipated when the energy broadening by scattering effects, which is estimated by magnitude of the longitudinal conductivity, exceeds the energy gap formed by the spin-orbit interaction~\cite{Shiomi2010}.
In contrast to the electric AHE where both longitudinal and transverse conductions are given by electrons, the thermal Hall conductivity in $\alpha$-RuCl$_3$ is carried by the itinerant Majorana fermions whereas the longitudinal thermal conducitivity is by phonons.
Thus, the scattering effects on Majorana fermions cannot be estimated from the magnitude of $\kappa_{xx}$.
Meanwhile, it has been pointed out that a large coupling between the Majorana fermions and the phonons is necessary to observe the quantized thermal Hall effect~\cite{VinklerAviv2018,Ye2018}.
We therefore speculate that the magnetic scattering effects on phonons, which broaden the Majorana energy bands through a spin-phonon coupling, may give rise to a closing of the Majorana gap protecting the chiral edge current.
The dominant magnetic scatterings on $\kappa_{xx}$~\cite{Hentrich2018} and the larger field effect on $\kappa_{xx}$ observed in the sample A also imply that magnetic impurities, rather than nonmagnetic ones, are more relevant to the dissipation mechanism of the Majorana gap.

The effects of disorders, such as bond randomness or vacancies, on the Kitaev model have been extensively studied in theory~\cite{Willans2010,Willans2011,ChuaFiete2011,Santhosh2012,AndradeVojta2014,Yamada2020,NasuMotome2020}.
Recently, it has been pointed out that $\kappa_{xx}$ is easily suppressed by disorders~\cite{NasuMotome2020} and that the quantized thermal Hall effect is dissipated by a closing of the Majorana gap~\cite{NasuMotome2020} or by the Anderson localization of the Majorana fermions~\cite{Yamada2020}.
These high sensitivities of $\kappa_{xx}$ and $\kappa_{xy}$ on disorders are consistent with our experimental results.
Clarifying further details of the disorder effects on the quantized thermal Hall effect by investigating the structure of the candidate materials or by artificially introducing disorders will be an important future issue.

In summary, we have investigated the sample dependence of the thermal Hall conductivity of the Kitaev candidate material $\alpha$-RuCl$_3$.
We confirm the reproducibility of the half-integer quantized thermal Hall effect in the sample with the largest longitudinal thermal conductivity.
We also find the magnitude of the longitudinal thermal conductivity is positively correlated to the field-induced increase of $\kappa_{xx}$ and the decrease of $\chi$ below $T_{\textrm N}$.
We suggest that suppressing the magnetic scattering on phonons is important to realize the quantized chiral edge current.

\begin{acknowledgments}
We thank Y. Kasahara, Y. Matsuda, and T. Shibauchi for fruitful discussions. This work was supported by KAKENHI (Grants-in-Aid for Scientific Research) Grant No. 17H01142, No. 19H00648, No. 19H01848, No.  19K03711 and No. 19K21842.
\end{acknowledgments}

	
%

\newpage

\renewcommand{\theequation}{S\arabic{equation}}
\setcounter{equation}{0}
\renewcommand{\thetable}{S\arabic{table}}
\setcounter{figure}{0}
\renewcommand{\thefigure}{S\arabic{figure}}

\begin{titlepage}
	\begin{center}
		\vspace*{12pt}
		{\Large Supplemental Material for ``Sample dependence of the half-integer quantized thermal Hall effect in the Kitaev candidate $\alpha$-RuCl$_3$"}
		\vspace{12pt} \\
		\begin{tabular}{c}
			M. Yamashita, J. Gouch, Y. Uwatoko, N. Kurita, H. Tanaka
		\end{tabular} \vspace{3pt} \\
	\end{center}
\end{titlepage}

\subsection{The temperature dependence of the heat capacity measured after the thermal-transport measurements}

\begin{figure}[!htb]
	\centering
	\includegraphics[width=0.8\linewidth]{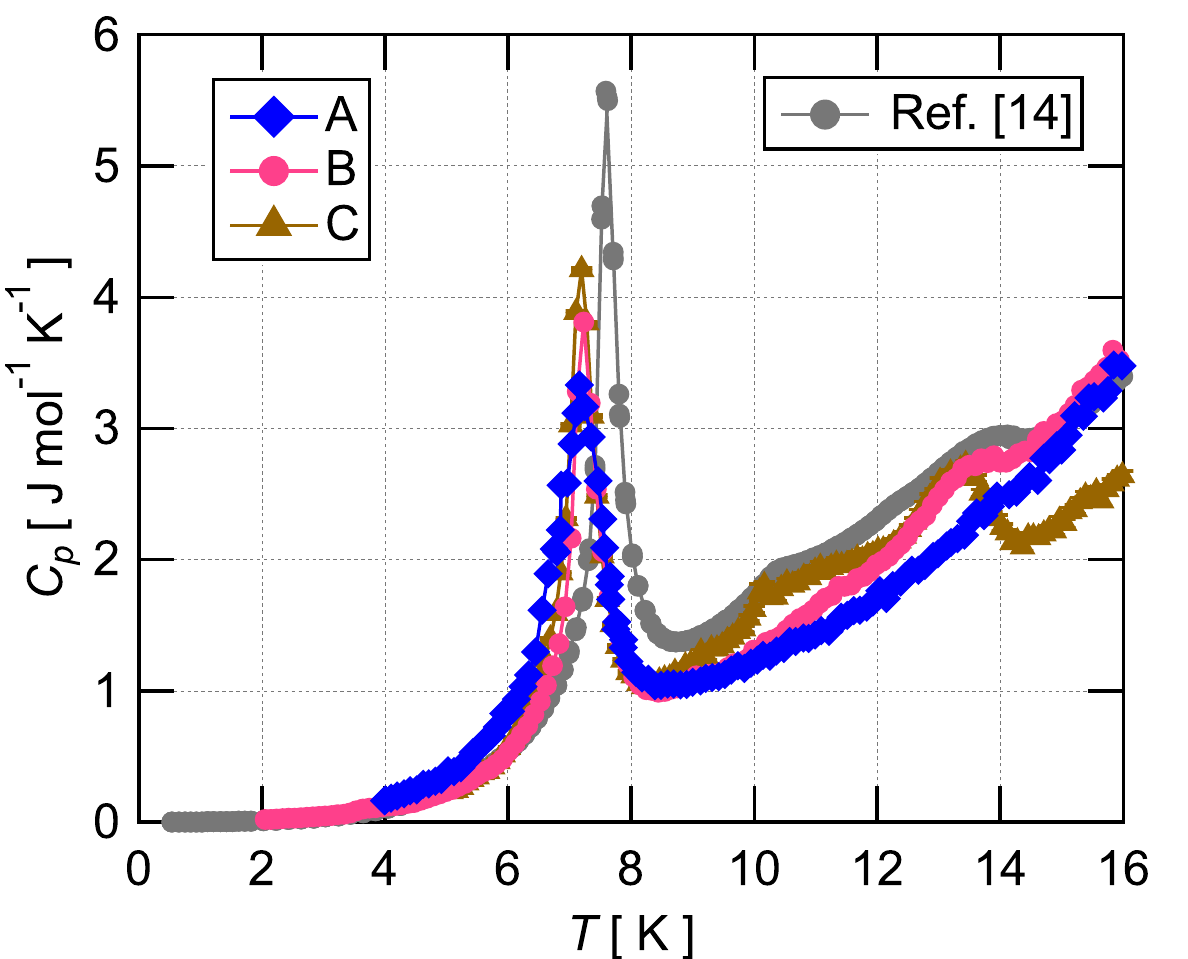}
	\caption{
		The temperature dependence of the heat capacity ($C_p$) of all samples at zero field.
		The data from Ref.\;\cite{Kubota2015} is shown for reference.
	}
	\label{C_p}
\end{figure}

To check the sample quality, we measured the temperature dependence of the heat capacity ($C_p$) at zero field after the thermal-transport measurements.
As shown in Fig.\,\ref{C_p}, $C_p$ of sample A shows no anomaly at 14\,K associated with stacking faults~\cite{Kubota2015}.
Although small anomalies at 10 and 14\,K are observed for sample B and C similar to those reported in Ref.\,\cite{Kubota2015}, the anomalies are much smaller than the peak at the antiferromagnetic ordering temperature.

Note that the absolute value of $C_p$ of our measurements has a relatively large error because of a large ambiguity in the mass estimation.
This is because we measured $C_p$ of samples with gold wires and pastes, which were used for thermal connections to the heaters and the thermometers, to avoid additional stresses on the sample to remove them.  
Despite this, we cannot exclude a possibility that additional stresses were applied on the sample when detaching the sample from the thermal-transport cell as the origin of the small anomalies observed in sample B and C.

\end{document}